%% file: uep_letter.tex
\documentclass[useAMS,usenatbib]{mn2e}
\usepackage{amssymb,amsmath,epsfig,times, natbib}


\title[A universal entropy profile for relaxed clusters?]{Galaxy cluster outskirts: a universal entropy profile for relaxed clusters?}
\author[S. A. Walker et al.]{S. A. Walker,$^1$\thanks{Email: 
    swalker@ast.cam.ac.uk} A. C. Fabian,$^1$ J. S. Sanders$^1$ and M. R. George$^2$ \\
  $^1$Institute of Astronomy, Madingley Road, Cambridge CB3 0HA \\
  $^2$Department of Astronomy, University of California, Berkeley, CA 94720, USA
}

\begin{document}

\maketitle

\begin{abstract}
 We fit a functional form for a universal ICM entropy profile to the scaled entropy profiles of a catalogue of X-ray galaxy cluster outskirts results, which are all relaxed cool core clusters at redshift below 0.25. We also investigate the functional form suggested by Lapi et al. and Cavaliere et al. for the behaviour of the entropy profile in the outskirts and find it to fit the data well outside 0.3$r_{200}$. We highlight the discrepancy in the entropy profile behaviour in the outskirts between observations and the numerical simulations of Burns et al., and show that the entropy profile flattening due to gas clumping calculated by Nagai \& Lau is insufficient to match observations, suggesting that gas clumping alone cannot be responsible for all of the entropy profile flattening in the cluster outskirts. The entropy profiles found with \emph{Suzaku} are found to be consistent with \emph{ROSAT}, \emph{XMM-Newton} and \emph{Planck} results. 
\end{abstract}


\section{Introduction}

 A breakthrough in the study of the low surface brightness X-ray emission from the intracluster medium (ICM) in the outskirts of galaxy clusters has been made possible by the low and stable particle background of \emph{Suzaku}, and a statistical sample of clusters studied to $r_{200}$\footnote{$r_{200}$ is the radius within which the mean mass density is 200 times the critical density for a flat universe.}  has started to emerge (see table \ref{Cluster_sample}). All of these are relaxed cool core clusters with redshifts of less than 0.25. Since these clusters have been chosen because they are very X-ray bright, they do not provide a representative sample of the actual cluster population, but nonetheless general trends have started to emerge. Understanding the thermodynamic properties of clusters in the outskirts is important for constraining the physical processes occurring in the ICM. We also need to understand the radius at which the hydrostatic equilibrium approximation breaks down, and the contributions of non-thermal pressure support which are expected to increase in the outskirts \citep{Lau2009} in order to calculate accurate masses out to $r_{200}$ and beyond.   
 
 One of the interesting findings from studies of the outskirts of galaxy clusters is that the scaled ICM entropy ($S=kT/n_{e}^{2/3}$) profile has a similar shape for each cluster, flattening off above $\sim 0.5r_{200}$ away from a powerlaw increase. The entropy profile is of fundamental importance because it both determines the structure of the ICM and provides a record of the thermodynamic history of the ICM. As described in \citet{Voit2005}, the entropy is more representative of the thermodynamic history of the ICM than temperature because when heated gas expands in a gravitational potential its thermal energy can be converted into gravitational potential energy. Introducing heat will always cause the entropy to rise while losing heat through radiative cooling will always cause the entropy to fall, whereas it is possible for the luminosity-weighted temperature to rise only slightly on the input of a large amount of energy, making it a less sensitive probe of ICM history \citep{Voit2002}.
 
  Here we compile a sample based on the papers summarised in table \ref{Cluster_sample}, and fit an analytic function to the scaled entropy profiles. In addition to the clusters studied with \emph{Suzaku}, we also add Abell 1835 and Abell 2204 which have been studied with Chandra out to $r_{100}$ and 0.7$r_{200}$ respectively. We also add the Virgo cluster which has been studied with XMM-Newton out to $r_{200}$ along one narrow strip. We do not include the clusters studied with \emph{Suzaku} in the outskirts described in \citet{AkamatsuRadioRelics} as these observations are aligned specifically with radio relics, and so are not representative of the ICM of the whole cluster. Compiling a sample like this reduces the effect of the small azimuthal coverage of some of the observations, allowing a more realistic average to be obtained which reduces the bias of looking along particular directions in a cluster which may not be representative of the cluster as a whole. For instance \citet{Eckert2012} has used ROSAT PSPC observations and found that clusters in the outskirts have significant ($\sim$ 70 percent) azimuthal scatter in their surface brightness when the clusters are divided into 12 sectors. Theoretically galaxy clusters are expected to be increasingly anisotropic in their outskirts with accretion occurring preferentially along large scale structure (LSS) filaments \citep{Burns2010}. 

\begin{table}
  \begin{center}
  \caption{Sample of galaxy cluster outskirts observations used.}
  \label{Cluster_sample}
  
    \leavevmode
    \begin{tabular}{llllll} \hline \hline
    Cluster &z& Reference & Plot symbol\\ \hline
Abell 2029 &0.0767 & \citet{Walker2012_A2029}a &Red square\\ 
    PKS 0745-191 & 0.1028& \citet{Walker2012_PKS0745}b &Grey square\\
    Abell 1795 &0.063 & \citet{Bautz2009} &Red triangle \\
   Abell 1413 & 0.143 & \citet{Hoshino2010} &Blue triangle\\
    Abell 2142 & 0.0899& \citet{Akamatsu2011} &Blue square\\
    Hydra A &0.0539 & \citet{Sato2012} &Cyan square\\
   Perseus & 0.0183& \citet{Simionescu2011} &Pink square\\
   Abell 1689 & 0.183& \citet{Kawaharada2010} &Red square\\
   Abell 1835 &0.253 & \citet{Bonamente2012} & Black square\\
   Abell 2204 &0.152 & \citet{Sanders2009} & Black triangle\\
   Virgo & 16.1 Mpc & \citet{Urban2011} & Green crosses\\
      \hline
    \end{tabular}
  \end{center}
\end{table}

As mentioned in \citet{Walker2012_PKS0745}b, the emerging picture for these low redshift, relaxed cool core clusters is that of an entropy profile which obeys the $r^{1.1}$ powerlaw (from \citealt{Voit2005}) from around 0.2$r_{200}$ out to around 0.5$r_{200}$, beyond which it flattens. The radius at which this flattening starts, and the extent of this flattening, provides an important test of numerical simulations of galaxy clusters, requiring an accurate understanding and modelling of the gravitational and non-gravitational heating mechanisms in the ICM. 

A number of physical processes are expected to affect the entropy profile in the outskirts. Gas clumping causes the gas density to be overestimated, thus causing the entropy to be underestimated. Gas clumping however is expected to be most significant outside $r_{200}$, \citep{Nagai2011}. 

A difference between the electron and ion temperatures in the ICM inside the accretion shock in the outskirts may also be important, and has been proposed in \citet{Hoshino2010} and \citet{Akamatsu2011} as a cause for entropy profile flattening. This would arise because the ions, which carry most of the kinetic energy of the accreting gas (due to their greater mass) are thermalised first after the accretion shock, and this energy is then transferred inefficiently and slowly to the electrons through electron-ion collisions. However the absence of a pressure drop in the outskirts of galaxy clusters in the Sunyaev-Zeldovich (SZ) effect observations with Planck \citep{PlanckV2012}, and the agreement between the Planck observations of the gas pressure with simulations suggests that the electrons and ions in the outskirts are in equilibrium. In addition, the hydrodynamical simulations of \citet{Wong2009} indicate that within $r_{100}$ the electron and ion temperatures are expected to differ by less than a percent. 

 \begin{figure*}
  \begin{center}
    \leavevmode
 
        \hbox{
      \includegraphics[width=87mm]{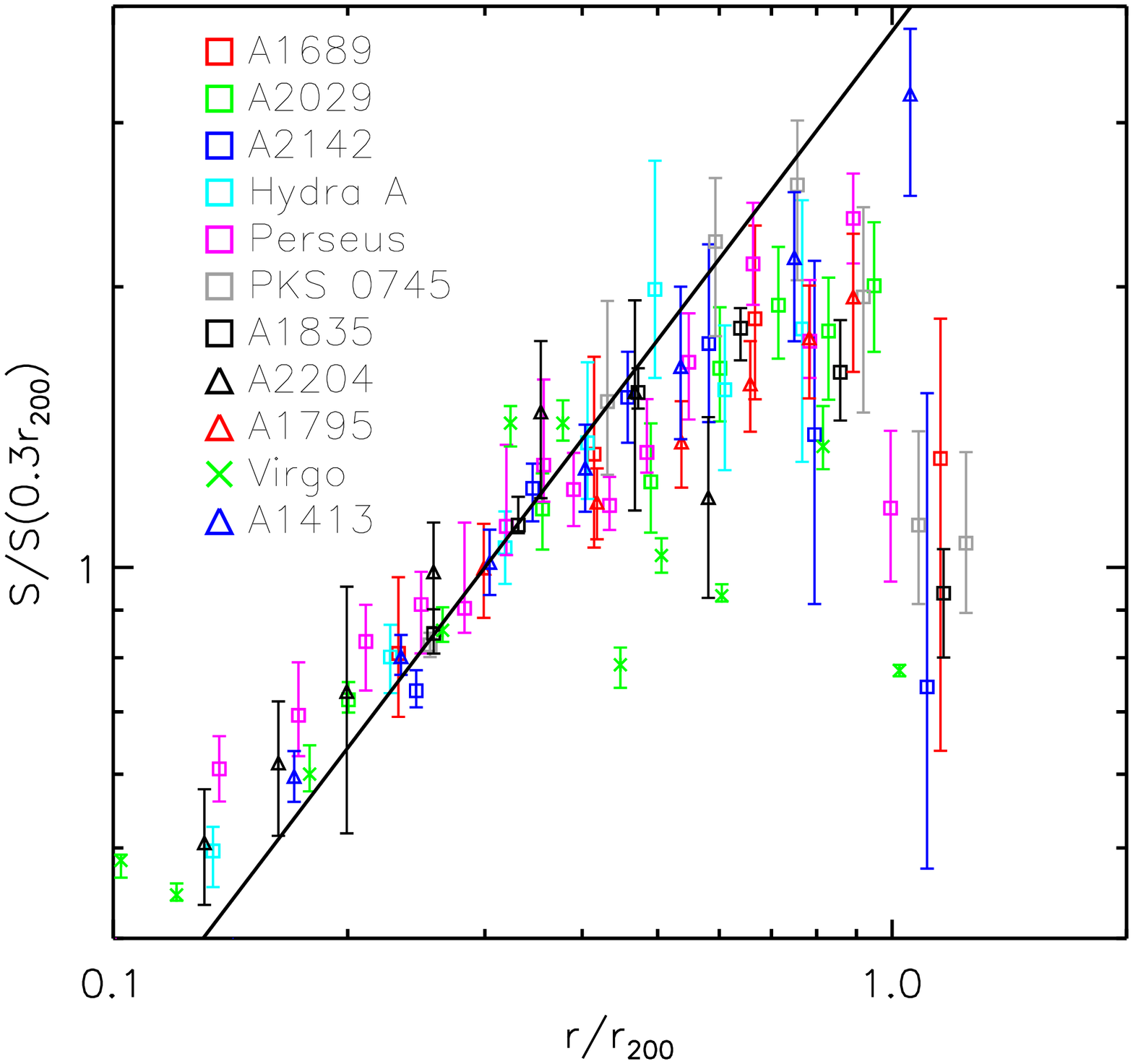}
            \includegraphics[width=87mm]{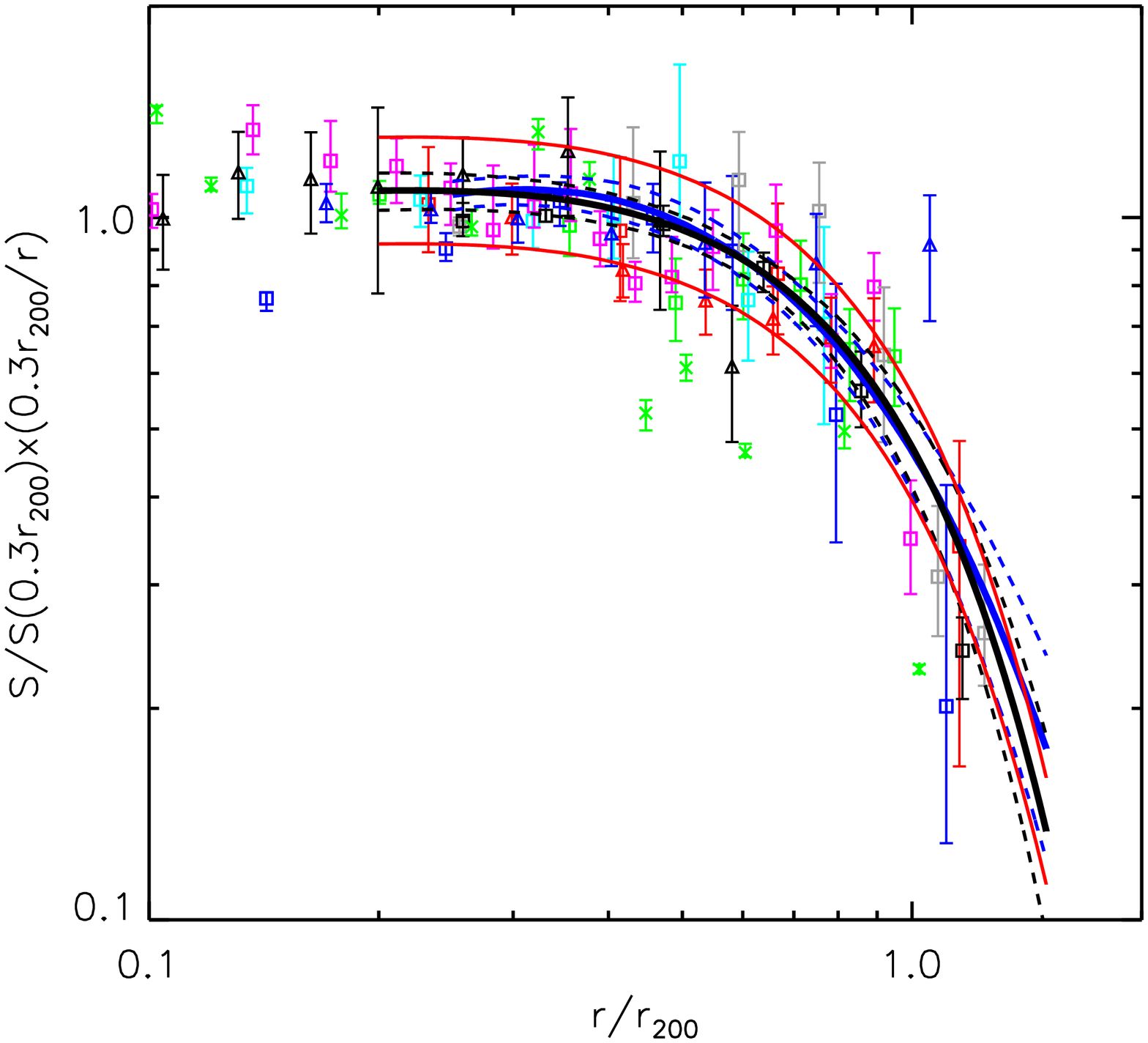}}
      \caption{\emph{Left}:Entropy profiles for the clusters shown in table \ref{Cluster_sample}, scaled by $S(0.3r_{200})$ . Individual clusters are colour coded as shown in table \ref{Cluster_sample}. The solid black line shows the $r^{1.1}$ powerlaw relation from \citet{Voit2005}. \emph{Right}:We plot $S(r)/r$ (scaled to 0.3$r_{200}$) to show the deviation from a powerlaw more clearly. The black line is the best fit line to the data outside $0.2 r_{200}$ using a form $S/S(0.3r_{200}) = A (r/r_{200})^{1.1} e^{-(r/Br_{200})^{2}}$. The best fit using the functional form of \citet{Cavaliere2011} (equation \ref{Cavaliere_eqn}) is shown by the blue line. For each model the 2 $\sigma$ variations calculated using Monte Carlo methods are shown by the dashed lines. The solid red lines show the range produced by density variations of 30 percent, which is the observed azimuthal density variation found near $r_{200}$ in \citet{Eckert2012}. }  
      \label{raw_data}
  \end{center}
\end{figure*}

\citet{Lapi2010} and \citet{Cavaliere2011} have proposed that entropy profile flattening in the outskirts of relaxed galaxy clusters is the natural result of the weakening of the accretion shock as it expands. This is a result of the accreting material originating from the tapering wings of the dark matter perturbation over a decreasing background density level due to the accelerated cosmic expansion.  Therefore the accreting gas falls through a progressively smaller potential drop as the accretion shock expands outwards. The weakening of the accretion shock as it expands reduces the entropy gain at the shock and also increases the amount of bulk energy passing across the shock by the infalling gas. This increases the level of turbulence and non-thermal pressure support in the outskirts, causing hydrostatic equilibrium mass estimates which use only the thermal pressure to underestimate the true mass (as has been found in \citealt{Lau2009}).

\citet{Lapi2010} and \citet{Cavaliere2011} propose a functional form for the entropy profile in the outskirts (outside $\sim R/3$) which can be written as:
\begin{eqnarray}
S/S(0.3r_{200}) = A_{C} (r/R)^{B_{C}} e^{C_{C}(1-(r/R))}
\label{Cavaliere_eqn} 
\end{eqnarray}
where $R$ is the virial radius, taken to be half the turnaround radius for the collapsing overdensity (which has detached from the Hubble flow) from which the cluster forms, which in a $\Lambda$CDM universe is $R=r_{100}$. This functional form has a maximum at $r/R=B_{C}/C_{C}$, and a gradient at $R$ which is $B_{C}-C_{C}$. \citet{Lapi2010} and \citet{Cavaliere2011} propose that as clusters get older the flattening and downturn of the entropy profile at large radius becomes more severe due to the reduction in the strength of the accretion shock. However the small redshift range covered by our sample (all have redshifts less than 0.25) means that no age evolution is discernible within our sample. The entropy profile flattening is not expected to occur in non-cool-core clusters, since for these younger clusters the high accretion rates and strong shocks should allow the entropy profile to follow the $r^{1.1}$ relation out to the virial radius, however their often disturbed morphologies complicates observations.  
 
 \section{Analysis}

 We use the entropy profiles for the clusters listed in table \ref{Cluster_sample} and scale them by their entropy at $0.3 r_{200} \simeq r_{2500}$, a region where the clusters all demonstrate a powerlaw increase in entropy and are in hydrostatic equilibrium since this is both far enough away from the core to avoid the affects of cooling and feedback processes, and  far enough from the outskirts to be unaffected by accretion processes. This is shown in Fig. \ref{raw_data} left. For the data from \citet{Bautz2009}, which presents analytic forms for the entropy profile instead of points, we show points centred on the annuli from which the spectra were extracted, and include only the data in the range in which the analytic forms for the temperature and density (and thus entropy) were fit (between 5$'$-17.5$'$). We do the same for the results of \citet{Bonamente2012}, in which a best fit analytic entropy profile line is presented rather than entropy data points for each annulus. 
 
\begin{figure}
  \begin{center}
    \leavevmode

      \includegraphics[width=87mm]{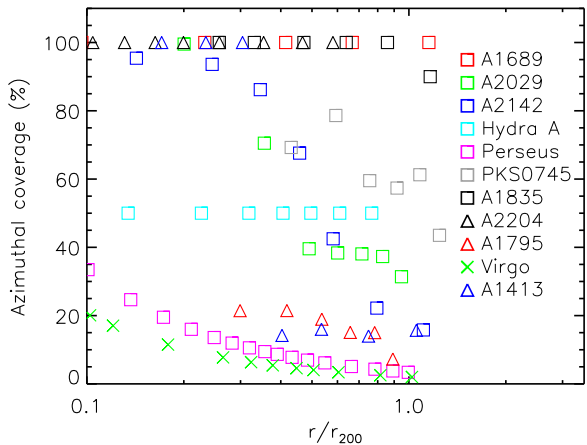}
      
      \caption{Percentage azimuthal coverage as a function of radius for the observations used. }  
      \label{az_coverage}
  \end{center}
\end{figure}

\begin{figure}
  \begin{center}
    \leavevmode
 
         \hbox{
      \includegraphics[width=80mm]{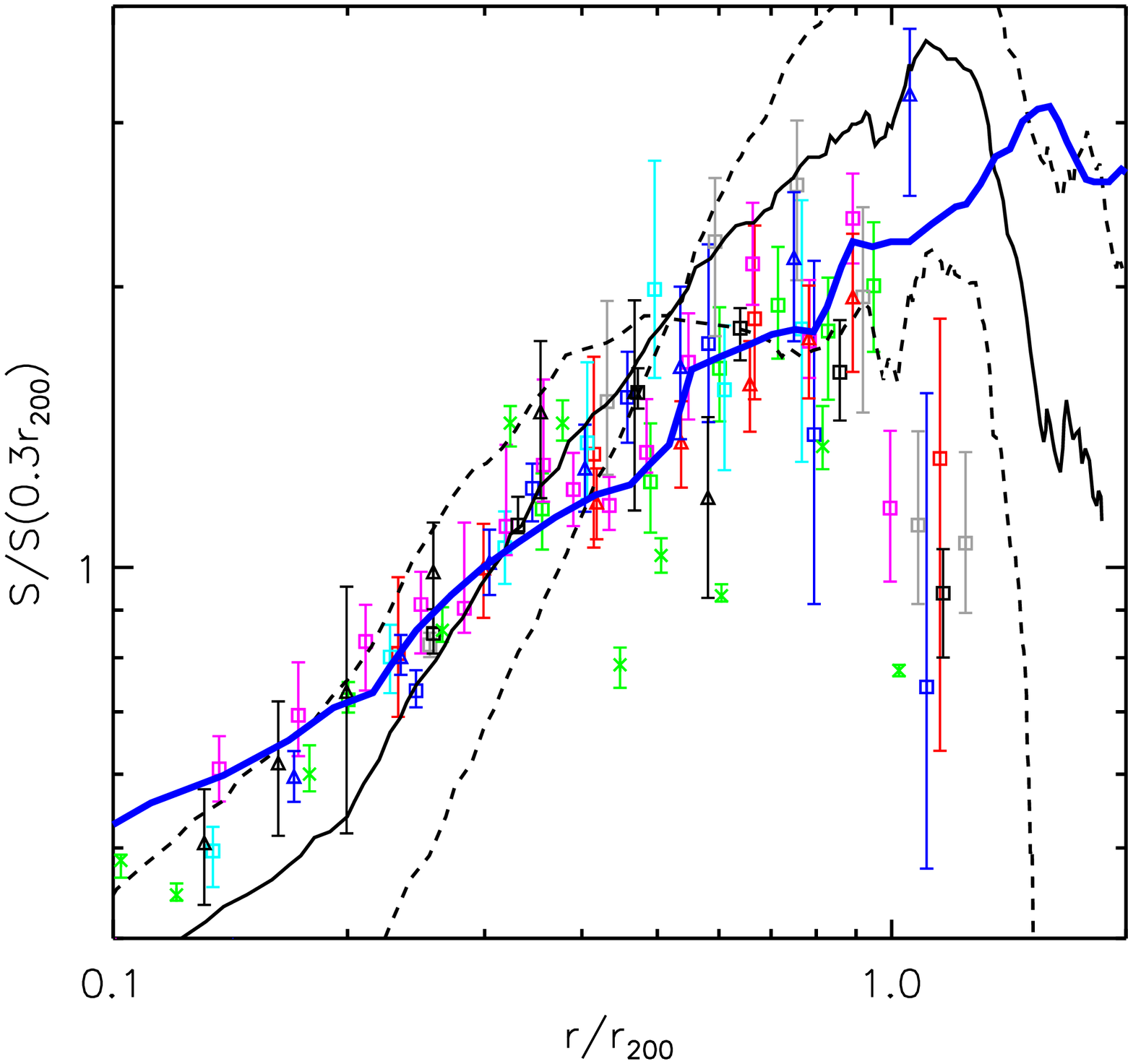}
            }          
      \caption{Same as Fig. \ref{raw_data} left but including the results of the simulations of \citet{Burns2010} as the black solid line. The dashed black lines show the standard deviation from the sample of numerical simulations studied in \citet{Burns2010} (which were normalised at $0.5 r_{200}$ for the sample, hence the zero scatter around this point). The blue solid line shows the results of the CSF simulations of \citet{Nagai2011} showing the flattening of the entropy profile as a result of gas clumping. In both cases the level of entropy flattening and turnover, and the radius at which the entropy starts to turnover, disagree with the observations.  }  
      \label{Burns_compare}
  \end{center}
\end{figure}

 The results for the Virgo cluster from \citet{Urban2011}, shown by the green crosses in Figs. \ref{raw_data} and \ref{Burns_compare}, are systematically below the general trend of the other clusters, indicating more severe entropy profile flattening, with significant oscillatory behaviour in the entropy profile with radius. This may be the result of a combination of the following factors. Virgo is the poorest and least massive cluster in our sample, which may indicate that entropy profile flattening is a function of cluster mass. The small azimuthal coverage of the Virgo observations (as shown in Fig. \ref{az_coverage}, at $r_{200}$ the azimuthal coverage is $\sim$ 2 percent) may mean that the results are not representative of the cluster as a whole, given the observed azimuthal variations in the surface brightness of the ICM in the outskirts of clusters \citep{Eckert2012}. For instance, if the observed strip happened to coincide with a region of higher than average gas clumping then this would anomalously give a larger inferred density, and thus a lower inferred entropy than the actual azimuthal average in the outskirts. The fact that there is a cold front at $\sim 0.2 r_{200}$ in the direction observed may also make the results unrepresentative of the whole cluster. 
 
 The outermost entropy measurement from Abell 1413 (blue triangles) also lies above the average relation. However as shown in Fig. \ref{az_coverage} the fraction of the ICM in the annulus which is studied is quite low ($\sim$ 15 percent), and so may not be representative of the whole cluster at that radius.  
 
 In Fig. \ref{raw_data} (right) we plot $S/r$ against r (scaling the profiles by $S(0.3r_{200})/0.3r_{200}$), which more clearly shows the deviation from a simple powerlaw above 0.5$r_{200}$. We find that the profile is fitted well by the functional form $S/S(0.3r_{200}) = A (r/r_{200})^{1.1} e^{-(r/Br_{200})^{2}}$ for $r \geq 0.2r_{200}$ with best fitting parameters $A=4.4^{+0.3}_{-0.1}$ and $B=1.0^{+0.03}_{-0.06}$, so that;
  \begin{eqnarray}
  S/S(0.3r_{200}) = 4.4 (r/r_{200})^{1.1} e^{-(r/r_{200})^{2}} 
  \label{expcutoff}
  \end{eqnarray}

  We also find the best fit to the scaled entropy profiles in the range $r \geq 0.3r_{200}$ using the functional form of equation \ref{Cavaliere_eqn} from \citet{Lapi2010} and \citet{Cavaliere2011}, which is found to model the entropy profiles well with best fit parameters $A_{C}=1.02^{+0.23}_{-0.08}$, $B_{C}=1.8^{+0.2}_{-0.2}$, $C_{C}=3.3^{+0.8}_{-0.2}$, so the best fit relation is
 \begin{eqnarray}
S/S(0.3r_{200}) = 1.02 (r/R)^{1.8} e^{3.3(1-(r/R))}
\label{Cavaliere_eqn2} 
\end{eqnarray}

  Since the errors on each parameter are correlated, the errors on the best fits were obtained by using a Monte Carlo method with 10000 trials, and the 2 $\sigma$ variations of the best fit models are shown by the dashed lines in Fig. \ref{raw_data} right. Black lines show equation \ref{expcutoff} while the blue lines show equation \ref{Cavaliere_eqn2}. When performing the fitting the entropy profiles from each cluster were also weighted by the azimuthal coverage of the observations of each cluster (shown in Fig. \ref{az_coverage}), so that more weight was given to observations with larger, more representative azimuthal coverage. This reduces the possible bias of observations which were taken along narrow strips which may not be representative of the cluster as a whole.  
  
 The solid red lines in Fig. \ref{raw_data} (right) show the effect of 30 percent density variations on the best fit entropy profile. This is the level of azimuthal scatter in the gas density inferred from the azimuthal scatter in the surface brightness of the clusters studied in \citet{Eckert2012} (where the observed surface brightness scatter was $\sim$ 70 percent around $r_{200}$). We find that the majority of the data lie within this range around the best fit profile, suggesting that most of the scatter around the best fit profile can be explained by the $\sim$30 percent azimuthal density variations found in \citet{Eckert2012}. The Virgo results are however inconsistent with the trend of the other clusters. This may be because the azimuthal scatter measured in \citet{Eckert2012} was found by dividing the clusters in their ROSAT sample into 12 sectors of opening angle 30 degrees, whereas the Virgo strip is much narrower than this (its opening angle is $\sim$ 8 degrees). It is therefore possible that the scatter measured in \citet{Eckert2012} underestimates the level of scatter at scales smaller than the sector size they used.
 
 In Fig. \ref{Burns_compare} (black lines) we compare the scaled entropy profiles with the similarly scaled predicted entropy profiles from the numerical simulations of \citet{Burns2010} obtained using the Eulerian adaptive mesh refinement (AMR) cosmology code \emph{Enzo} using only adiabatic gas physics (thus neglecting non-gravitational heating, cooling, magnetic fields, and cosmic rays). We find that the radius at which entropy flattening begins is higher in the simulations of \citet{Burns2010} than in the observations, suggesting that more physical processes need to be included in the numerical simulations to accurately describe the observed properties of clusters. The simulations of \citet{Kravtsov2012} also show the entropy profile flattening starting outside $r_{100}$, at a higher radius than is seen in the observations. A possibility is that the simulations may overestimate the number of major mergers at late cosmic times, causing the strength of the accretion shocks to be overestimated and increasing the radius at which the entropy profile flattening begins.   
 
In Fig. \ref{Burns_compare} we also compare the observed entropy profiles with the profile shape obtained in the gas clumping simulations of \citet{Nagai2011} (solid blue line), which included the effects of cooling and star formation (CSF). The clumping effect only becomes significant outside $r_{200}$, and is not large enough to account for the amount of entropy flattening and turnover. 

\begin{figure}
  \begin{center}
    \leavevmode
 
      \vbox{
      \includegraphics[width=80mm]{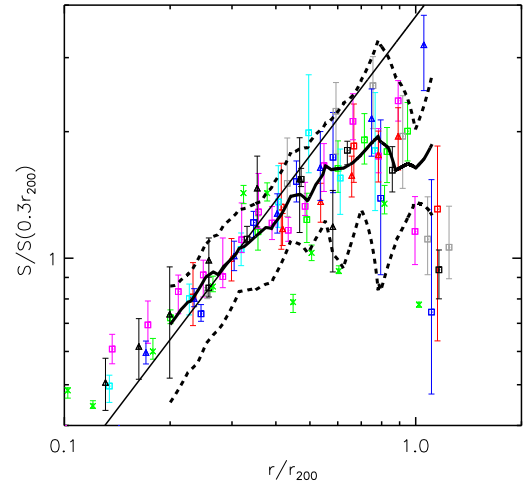}
            \includegraphics[width=80mm]{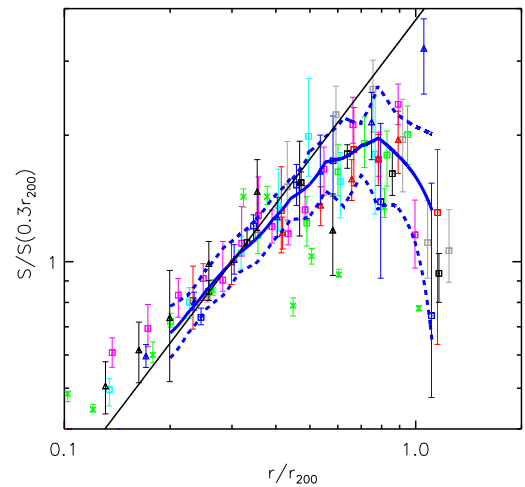}}

      \caption{\emph{Top}:The thick solid black line shows the entropy profile obtained by combining the pressure profile of \citet{PlanckV2012} (obtained from a sample of 62 clusters) and the density profile obtained with ROSAT PSPC in \citet{Eckert2012} (obtained from a sample of 31 clusters). The dashed thick black lines show the scatter. \emph{Bottom}: The solid blue line shows the entropy profile obtained by combining the ROSAT density profile from \citet{Eckert2012} with the extrapolated XMM-Newton temperature profile shown in Fig. \ref{XMM_temps}, and the 1-$\sigma$ error is shown by the thick blue dashed lines. In both plots the thin black line shows the $r^{1.1}$ powerlaw relation.   }  
      \label{comparetoothers}
  \end{center}
\end{figure}

 \section{Comparing to ROSAT, XMM-Newton and Planck results}
 
 \citet{PlanckV2012} obtained a pressure profile for a stacked sample of 62 clusters using the SZ-effect reaching out to $3r_{500} \simeq 2r_{200}$, while \cite{Eckert2012} obtained an average density profile from a sample of 31 clusters obtained with \emph{ROSAT} PSPC observations out to $r_{200}$. Using that $S \propto P/n_{e}^{5/3}$ we can combine these to obtain a scaled entropy profile, and this is plotted in Fig. \ref{comparetoothers} (top) as the thick black line. The resulting entropy profile agrees well with the \emph{Suzaku} observations, flattening off above 0.5$r_{200}$. 
  
\begin{figure}
  \begin{center}
    \leavevmode
      
      \includegraphics[width=78mm]{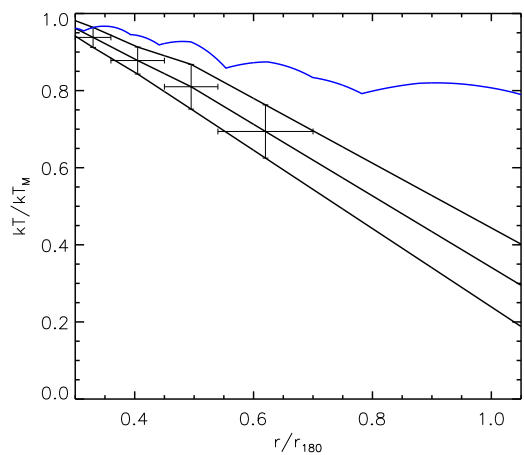}
      
      \caption{Black crosses show the mean temperature profile (scaled by the mean cluster temperature $kT_{M}$) obtained in \citet{Leccardi2008} with XMM-Newton for a sample of 50 clusters out to $\sim$0.7$r_{200}$ (their figure 20). The solid black lines show linear extrapolations of this profile out to $r_{200}$. The solid blue line shows the temperature profile that would be needed to obtain an $r^{1.1}$ entropy relation when the density profile obtained with ROSAT is used. }  
      \label{XMM_temps}
  \end{center}
\end{figure}

 As another test we use the average temperature profile found in \citet{Leccardi2008} out to $\sim$0.7$r_{200}$ using \emph{XMM-Newton} observations of a sample of $\approx$ 50 clusters, and extrapolate this out linearly to $r_{200}$, as shown in Fig. \ref{XMM_temps}. When combined with the \emph{ROSAT} density profile from \citet{Eckert2012} the resulting scaled entropy profile is shown by the solid blue line in Fig. \ref{comparetoothers} (bottom), which again agrees well with the \emph{Suzaku} results. 
 
 In order for the entropy profile to follow a $r^{1.1}$ relation out to $r_{200}$ and beyond, the temperature profile would need to be the blue line shown in Fig. \ref{XMM_temps} when the density profile from \citep{Eckert2012} is used, which shows a much shallower temperature decline in the outskirts than is observed. 
 
\section{Summary}
We have compiled a sample of relaxed cool core galaxy clusters (with $z \leq 0.25$) whose ICM has been studied to  $\sim r_{200}$. We find that the scaled entropy profiles are best fit using a function of the form $S/S(0.3r_{200}) = 4.4 (r/r_{200})^{1.1} e^{-(r/r_{200})^{2}}$. We also investigate the analytic function proposed by \citet{Cavaliere2011}, and have found it to fit the data well outside 0.3$r_{200}$ (equation \ref{Cavaliere_eqn2}). This adds support to the suggestion of \citet{Lapi2010} and \citet{Cavaliere2011} that the flattening and downturn of the ICM entropy in the outskirts is the result of the decreasing strength of the accretion shock as it expands, due to the accreting matter coming from the tapering wings of the dark matter perturbation. The increase in bulk energy seeping through the shock is expected to lead to turbulence and non-thermal pressure support, which are predictions that could be tested by next generation instruments studying cluster outskirts. 

The observed scaled entropy profiles turnover at a radius lower than that predicted by the numerical simulations of \citet{Burns2010} which used only adiabatic gas physics. The entropy profile flattening due to gas clumping calculated in the CSF numerical simulations of \citet{Nagai2011} is insufficient to match the observations (Fig. \ref{Burns_compare}), suggesting that clumping alone cannot be responsible for the flattening and turnover of the entropy profile, indicating that some other process must contribute. Clumping is required to explain the observed excess of the gas mass fraction over the mean cosmic baryon fraction around $r_{200}$ (\citealt{Simionescu2011}; \citealt{Walker2012_PKS0745}b; \citealt{PlanckV2012}), though part of this excess may be the result of increasing non-thermal pressure support in the outskirts causing the total mass estimates using only thermal pressure to be underestimated by around 20 percent at $r_{200}$ (\citealt{Lau2009}; \citealt{Walker2012_PKS0745}b). 

The \emph{Suzaku} entropy profiles are in strong agreement with the entropy profiles obtained by combining the mean pressure profile obtained with \emph{Planck} \citep{PlanckV2012}, the mean density profile obtained with \emph{ROSAT} PSPC \citep{Eckert2012} and the extrapolation of the mean temperature profile obtained with \emph{XMM-Newton} \citep{Leccardi2008}.

\section*{Acknowledgements}
SAW is supported by STFC, and ACF thanks the Royal Society. MRG acknowledges
support from an NSF Graduate Research Fellowship and NASA grant NNX10AR49G. This research has used data from the $Suzaku$
telescope, a joint mission between JAXA and NASA.

\bibliographystyle{mn2e}
\bibliography{uep_letter}

\end{document}